\documentclass[prd, onecolumn, floatfix, letterpaper, nofootinbib, amsmath, amssymb, superscriptaddress]{revtex4}
\usepackage{caption}
\usepackage{graphicx}
\usepackage{epsfig}
\usepackage{bm}
\usepackage{amsfonts}
\usepackage{comment}
\usepackage{tikz}
\usetikzlibrary{shapes,arrows}
\usepackage{subfigure}

\usepackage{float}

\usepackage{hyperref}

\usepackage{pdfpages}

\usepackage{hyperref}

\usepackage{amsmath}

\usepackage{color}

\newbox\pippobox

\def\be{\begin{equation}}
\def\ee{\end{equation}}
\def\ba{\begin{eqnarray}}
\def\ea{\end{eqnarray}}
\newcommand {\lla} {\ {\raise-.5ex\hbox{$\buildrel<\over\sim$}}\ }

\usepackage[T1]{fontenc}
\usepackage[latin1]{inputenc}
\usepackage{graphicx}
\usepackage[english]{babel}
\usepackage{amsmath}
\usepackage{amssymb}
\usepackage{amsfonts}

\def\be{\begin{equation}}
\def\ee{\end{equation}}
\def\bea{\begin{eqnarray}}
\def\eea{\end{eqnarray}}

\def\spose#1{\hbox to 0pt{#1\hss}}

\def\lta{\mathrel{\spose{\lower 3pt\hbox{$\mathchar"218$}}
     \raise 2.0pt\hbox{$\mathchar"13C$}}}
\def\gta{\mathrel{\spose{\lower 3pt\hbox{$\mathchar"218$}}
     \raise 2.0pt\hbox{$\mathchar"13E$}}}

\usepackage{mciteplus}

\begin{document}

\title{Preheating and Entropy Perturbations in Axion Monodromy Inflation}

\author{Evan McDonough}
\email{ evanmc@physics.mcgill.ca}
\affiliation{Department of Physics, McGill University, Montr\'eal, QC H3A 2T8, Canada}

\author{Hossein Bazrafshan Moghaddam}
\email{bazrafshan@physics.mcgill.ca}
\affiliation{Department of Physics, McGill University, Montr\'eal, QC H3A 2T8, Canada}

\author{Robert H. Brandenberger}
\email{rhb@physics.mcgill.ca}
\affiliation{Department of Physics, McGill University, Montr\'eal, QC H3A 2T8, Canada, and
Institute for Theoretical Studies, ETH Z\"urich, CH-8092 Z\"urich, Switzerland}

\begin{abstract}      

We study the preheating of gauge fields in a simple axion monodromy model and
compute the induced entropy perturbations and their effect on the curvature fluctuations.
We find that the correction to the spectrum of curvature perturbations has a blue spectrum with
index $n_s = 5/2$. Hence, these induced modes are harmless for the observed
structure of the universe. Since the spectrum is blue, there is the danger of overproduction
of primordial black holes. However, we show that the observational constraints are
easily satisfied.

\end{abstract}

\maketitle

\section{Introduction}

Axion monodromy inflation \cite{Liam} (see also \cite{Eva})
may be the most promising way to obtain large field
inflation in the context of superstring theory \footnote{Note, however, that there may be
constraints on the scenario coming from string back-reaction effects \cite{Conlon}
and from the ``Weak Gravity Conjecture" \cite{WGC}.}. Large field inflation models
have the advantage over most small field models in that the inflationary slow-roll
trajectory is a local attractor in initial condition space \cite{attractor} (see e.g. \cite{RHB_IC} for
a recent review of this issue). 

Axion monodromy models contain, in addition to
the axion field (which plays the role of the inflaton), gauge fields to which the axion
couples via a Pontryagin term in the effective action. As a consequence, during
the post-inflationary phase when the inflaton starts to oscillate, there is a preheating
type instability in the gauge field equation of motion, and there can
be explosive gauge field particle production. This, at second order in the amplitude
of the gauge field perturbations, induces a growing curvature fluctuation mode.

The amplitude of the induced curvature fluctuations is constrained by observations.
On one hand, on cosmological scales the amplitude of the induced curvature
fluctuations must be smaller than the observed perturbations \footnote{They must
be strictly smaller since the observed fluctuations are well described by a Gaussian
process, whereas the induced fluctuations due to the gauge field perturbations
have non-Gaussian statistics.}. This is easy to satisfy if the spectrum of the
induced fluctuations is blue, as in our case. If the spectrum is blue then, on the
other hand, we must worry about the possible over-production of primordial
black holes. 

In this paper we will show that both sets of constraints are satisfied. We will
first study the preheating of the gauge field fluctuations. Then, we compute
the resulting entropy fluctuations and determine the induced curvature
fluctuations. We find that the power spectrum of these perturbations is
deeply blue, with a spectral index of $n_s = 4$, which gives a leading correction to the 
curvature power spectrum with index $n_s = 5/2$. Hence, these perturbations
have a completely negligible effect on cosmological scales. The amplitude
of the perturbations on small scales is influenced both by the tachyonic
growth of the modes during inflation, and by the instability during the
preheating phase. However, for parameter values used in axion monodromy
models, we find that the constraints from possible over-abundance of
primordial black holes are easily satisfied.

The Pontryagin term which couples the axion to a gauge field and which is playing 
the key role in our study has been studied in detail in recent years. The tachyonic
instability which the gauge field experiences in the presence of a rolling axion field
during inflation has been investigated in \cite{Anber}. The amplified gauge fields, in
turn, lead to axion fluctuations which induce non-Gaussianities in the adiabatic
curvature perturbations \cite{Barnaby}.  It was realized that the spectrum of these
fluctuations is blue, and there hence are potential constraints on the theory due
to possible over-production of primordial black holes. This has been studied e.g.
in \cite{Bugaev, Pajer} (see also \cite{Lin}). The Pontryagin term in the joint action
of the axion and gauge field can lead to overdamped motion of the axion in
which the axion field value is set by the coupling to the gauge field, and both
the acceleration term and the velocity term in the axion equation of motion are
negligible \cite{Anber}. This can lead to inflation on steep potentials for
sub-Planckian field values \cite{Anber, Wyman}. The same Pontryagin term
also leads to an inverse cascade \cite{Juerg} for cosmological magnetic fields,
and it can be used to provide a scaling quintessence model for dark energy \cite{Juerg2}.

\section{Review of Axion Monodromy Inflation}

Axions are ubiquitous in string theory \cite{GSW, Witten}. They arise
in string compactifications by integrating gauge potentials over
non-trivial cycles of the compactification. In the absense of branes,
the potential for the axions is classically flat, and obtains periodic
terms from non-perturbative effects. However, in the presence of
branes the periodicity of the axion potential is broken. The axion
acquires an infinite field range with a potential which is slowly
rising as the absolute value of the field increases. In the original
example studied in \cite{Liam}, the potential is linear at large
field values.

A `realistic' construction of Axion Monodromy Inflation requires three distinct 
sectors: (1) the monodromy brane, (2) moduli stabilization, and (3) a 
realization of the Standard Model. Typically these each are achieved 
via a brane construction: the DBI action of a D5 brane induces monodromy 
for the axion associated with the NS two-form (the axion we will
focus our attention on), a stack of D7 branes induces gaugino condensation 
which fixes the radial modulus of the internal space, and a set of intersecting 
branes realizes the (extension of the) Standard Model. Each of these sectors comes 
with its own gauge theory: the monodromy brane has a U(1)
Super-Yang-Mills (SYM) theory, the stack of D7 branes has a SU(N) SYM, while the 
intersecting branes have either a GUT group (e.g. SU(5)) or the Standard Model 
group SU(3)$\times$SU(2)$\times$U(1). The axion of axion monodromy inflation 
is a bulk field, and thus couples to and can lose energy to each sector. There
may be phenomenological issues which arise when the energy loss of
the axion into other sectors is considered, but we will not address this
issue here.

In this work we will focus on a minimal setup of axion monodromy inflation in
which we only consider the monodromy brane and its associated $U(1)$
gauge field. For the case of the $B_2$ axion $\phi$, this gives the following 4d action
\footnote{There is additionally a coupling $\phi$ to $F_{\mu \nu} F^{\mu \nu}$, of the form
\begin{equation}
\frac{\beta}{\Sigma} V(\phi) \; F_{\mu \nu} F^{\mu \nu} ,
\end{equation}
which comes from the $(\alpha ')^2$ correction to the DBI action. The coupling 
constant of this term is smaller than the coupling to $F \tilde{F}$ by a factor of
 $C_0$, where $C_0\sim 10^2$ in cosmological models based on compactifications 
which stabilize moduli via gaugino condensation on D7 branes 
(see e.g. \cite{Rummel:2011cd}). We will ignore 
 the effect of this term in the current work.}  (in (-,+,+,+) signature)
\begin{equation}
\label{axionaction}
\mathcal{L} \, = \,  - (1/2) (\partial \phi)^2 + V(\phi) -  \frac{1}{4} F_{\mu \nu} F^{\mu \nu} 
+ \frac{1}{\Lambda}\; \phi\,  \; F_{\mu \nu} \widetilde{F}^{\mu \nu} ,
\end{equation}
where $\Lambda$ is a UV scale, and is different from the axion decay constant. 
The potential $V(\phi)$ is the monodromy potential:
\begin{equation}
\label{monopotential}
V(\phi) \, = \,  \mu^{3} \sqrt{\phi_c ^2 + \phi^2 } \, ,
\end{equation}
where $\mu$ is an energy scale whose value can be determined
from the observed magnitude of the cosmic microwave background
anisotropies, and where $\phi_c < m_{pl}$ is a constant, $m_{pl}$
denoting the Planck mass. The field strength 
\be
F_{\mu \nu} \, = \, \partial_\mu A_{\nu } - \partial_\nu A_{\mu}
\ee
is that of the abelian gauge field which lives on the brane world-volume, and we have 
neglected fermions, as preheating into fermions is inefficient. String models of axion 
monodromy and the resulting values of $\Lambda$ are discussed in Appendix A 
(see also \cite{momo}). From the point of view of effective field theory, we would 
expect $\Lambda$ to be given either by the string scale or the Planck scale. 
More stringent, though indirect, constraints come from models of early universe 
cosmology based upon this coupling. The gaussianity of the CMB constrains the parameter $\xi$, 
which we will define shortly, to be $\xi_{*} \lesssim 2.2$ at the moment when the pivot scale $k_{*}$ 
exits the horizon \cite{Ade:2015lrj}, see also \cite{Barnaby, Meerburg}, which can be translated to a 
bound $\Lambda^{-1} \leq 12 \, m_{pl}$. Recent results on the validity of perturbation theory during inflation 
\cite{Ferreira:2015omg} constrain $\xi$ to be $\xi \leq 3.5$, which correspond to an even tighter 
constraint on $\xi_{\ast}$ (if the whole inflationary trajectory is to be treated perturbatively). 
Given these considerations we will take a conservative approach,  
and work with an upper bound $\Lambda^{-1} \leq \mathcal{O}(1) m_{pl} \,^{-1} $. 

%

\section{Background Evolution}

We assume that the axion starts out in the large field region $\phi \gg m_{pl}$
where the slow-roll approximation
\be
3 H {\dot \phi} \, = \, - V^{\prime}(\phi) \, \simeq \, - \mu^3
\ee
of the equation of motion is self-consistent. The end of inflation occurs
at the field value when the slow-roll approximation breaks down, at which point $(1/2) \dot{\phi}^2=V$. This
takes place when 
\be
|\phi| \, \equiv \, \phi_e \, = \, \frac{1}{ \sqrt{6}} m_{pl} \, ,
\ee
and the kinetic energy at this point is
\be \label{velend}
\frac{1}{2} {\dot \phi}^2_{\phi  = \phi_e} \, = \, \frac{1}{\sqrt{6}} \mu^3 m_{pl} \, .
\ee
The value of the Hubble constant at the end of inflation is $H = H_{e}$ with
\be \label{Hend}
H_{e} \, = \, 2^{-1/4} 3^{-3/4} m_{pl}^{-1/2} \mu^{3/2} \, .
\ee
After inflation  ends $\phi$ begins anharmonic motion about the
ground state $\phi = 0$. As long as we can neglect the expansion of
space and the loss of energy by particle production, the motion is
periodic but anharmonic.

The value of $\mu$ is set by the observed amplitude of the cosmic microwave
background (CMB) anisotropies. A simple application of the usual
theory of cosmological perturbations (see e.g. \cite{MFB} for a
detailed review, and \cite{RHBfluctrev} for an overview) shows that the
power spectrum ${\cal P}_{\zeta}$ of the primordial\footnote{We add the word
``primordial'' to make a distinction between the original fluctuations and the
induced ones which will be the focus of this paper.} curvature fluctuation
$\zeta$ has the amplitude
\be
{\cal P}_{\zeta} \, \sim \bigl( \frac{\mu}{m_{pl}} \bigr)^3 ,
\ee
from which it follows that
\be
\mu \, \sim \, 6 \times 10^{-4} m_{pl} \, .
\ee
 
\section{Preheating of Gauge Field Fluctuations}

As first pointed out in \cite{TB} and \cite{DK}, a periodic axion background can
lead to explosive particle production for all fields coupled to the axion. This effect
is called ``preheating'' \cite{KLS, STB, KLS2} (see also \cite{ABCM, Karouby} for
reviews). Here we will consider the resonance of the gauge field fluctuations 
\footnote{There is the also a possibility that there is an efficient self-resonance of 
the inflaton, leading to oscillons \cite{oscillons}.  Oscillon formation occurs once 
the amplitude of $\phi$ oscillations falls below $\phi_c$, as defined in equation 
(\ref{monopotential}). Provided that $\phi_c$ is small compared to the initial 
amplitude of oscillations, which is indeed the case in realistic string embeddings, 
oscillon formation will not occur until preheating in to gauge fields has ceased 
to be efficient, and will not occur at all if preheating into gauge fields is efficient 
enough to halt the oscillatory motion of $\phi$. Given this, we will not consider 
oscillon formation in this work, although this does deserve further attention.}

The equation of motion for the linear fluctuations of $A_{\mu}$ is
(see e.g. \cite{Anber, Giblin, Barnaby})
\be \label{EoMA}
\frac{d^2 {A}_{k\pm}}{d \tau^2} + \left( k^2 \pm 2k \frac{\xi}{\tau} \right) A_{k\pm} \, = \, 0 \, , 
\ee
where $\pm$ denote the two polarizations of the gauge field, $\tau$ is conformal time,
$k$ indicates a comoving mode, and $\xi$ is given by\footnote{Our definition of $\xi$ is 
equivalent to the definition used in \cite{Anber, Giblin, Barnaby} with the identification 
$\alpha/f=4 / \Lambda$.}
\begin{equation} 
\xi \, = \, \frac{2 \dot{\phi}}{\Lambda H} \, ,
\end{equation}
where $H$ is the Hubble expansion rate and $\phi$ is the background field, and an
overdot denotes the derivative with respect to physical time. As long as the slow-roll
approximation is valid, $\xi$ can be taken to be constant. This is the equation relevant
during the inflationary period. 

As Eq. (\ref{EoMA}) shows, for one of the polarization states 
there is a tachyonic instability (see e.g. \cite{tachyonic}
for an initial discussion of tachyonic instabilities in reheating) already during inflation for
long wavelength modes, i.e. modes which obey
\be
k  - \frac{2 \xi}{|\tau|} = k - \frac{4|\dot{\phi}_I|}{\Lambda H |\tau|} \, < \, 0 \, ,
\ee
where the subscript $I$ indicates that the time derivative is evaluated during slow-roll
inflation. The critical wavelength beyond which there is a tachyonic instability
has a fixed value in physical coordinates if we take $H$ and ${\dot \phi}$ to
be constant in time. The critical wavelength can be called a ``gauge horizon''
and it plays a similar role as the Hubble radius (Hubble horizon) for cosmological
perturbations. The gauge horizon is proportional to the Hubble radius, its
physical wavenumber $k_p$ being given by
\be
k_p \, = \, 2 \xi H \, .
\ee

For modes which start in their vacuum state deep inside the horizon,
the tachyonic resonance \cite{tachyonic} leads to squeezing of the mode function. The
Floquet exponent is proportional to $k$, and hence, among all the modes which
become super-horizon (meaning super-gauge horizon) by the end of inflation, the ones
which undergo the most squeezing are the ones which exit shortly before the end
of inflation, i.e. whose  comoving wavenumbers is given by
\be 
k \, = \, k_{*} \, \equiv \,  2 \xi H \, ,
\ee
if we normalize the cosmological scale factor to be $a(t) = 1$ at the end of
inflation. The value of $k_{*}$ is determined by the Hubble rate and the
axion field velocity at the end of the period of inflation.

It can be shown \cite{Anber} that the mode function prepared by inflation is
\bea
\label{modefunctioninflation}
{A_{k+} ^{(0)}}  \, &=& \, 
\frac{2^{-1/4}}{\sqrt{2k}} \left( \frac{k}{ \xi a H}\right)^{1/4} e^{\pi \xi - 4 \xi \sqrt{ k / 2 \xi a H}} \nonumber \\
{A_{k-} ^{(0)}} \, &=& \, 0 ,
\eea
where $+/-$ denote the positive/negative chirality mode (the $-$ mode is not amplified during inflation).
This corresponds to a highly blue spectrum of gauge field fluctuations with an ultraviolet cutoff
which is set by the gauge horizon; the cutoff comes from the second term in the exponential
on the right hand side of (\ref{modefunctioninflation}). The major amplification factor $F_I$ of the amplitude
is
\be \label{ampl1}
F_I \, = \, e^{\pi \xi} \, .
\ee

For the specific potential (\ref{monopotential}) of axion monodromy inflation the values of $k_{*}$ and $\xi$ are
(making use of (\ref{velend}) and (\ref{Hend}) )
\bea
k_{*} \, &=& \, 4 \bigl( \frac{2}{3} \bigr)^{1/4} m_{pl}^{1/2} \mu^{3/2} \Lambda^{-1} \label{kstar} \\
\xi \, &=& \, 2\sqrt{6} \frac{m_{pl}}{\Lambda} \, . \label{xieq}
\eea
This shows that if $\Lambda \ll m_{pl}$ there is a large enhancement of the amplitude of
$A_k$ during inflation. On the other hand, if $\Lambda \gg m_{pl}$, then the growth is
negligible. For small values of $\Lambda$ (i.e. large values of $\xi$), the ``gauge horizon''
is smaller than the Hubble horizon, whereas for large values of $\Lambda$ the opposite
is true.

As mentioned above, the power spectrum ${\cal P}_{A}$ of gauge field fluctuations is blue. 
On length scales larger that the gauge horizon we have
\be
{\cal P}_{A}(k) \, \equiv \, k^3 |A_k|^2 \, \sim \, k^{5/2} \, .
\ee 

During reheating the expansion of space can be
neglected \cite{TB} and  the equation (\ref{EoMA}) becomes
\begin{equation}
\label{AEOM}
{\ddot{A}_{k\pm}} + \left( k^2 \pm  4\frac{k}{\Lambda} {\dot \phi} \right) A_{k\pm} \, = \, 0 .
\end{equation}
We immediately see that the tachyonic resonance which was present during
the period of inflation persists during the preheating period when $\phi$ undergoes
damped anharmonic oscillations about $\phi = 0$. While ${\dot \phi}$ is negative,
then the same polarization mode gets amplified as during inflation. During the
second half cycle, when ${\dot \phi} > 0$, it is the other mode which is amplified
while the original mode oscillates.

To obtain an order of magnitude estimate of the amplification of $A_k$ during preheating,
we focus on the first oscillation period (when the Floquet exponent of the
instability is largest). We focus on the first quarter of the oscillation period $T$
when $\phi$ is decreasing from $\phi = \phi_e$ to $\phi = 0$. The velocity
during most of this time interval is approximately ${\dot \phi}_e$ (see (\ref{velend})).
The amplitude of $A_k$ grows exponentially at a rate (for $k/k_{*}<1$), 
\be
\mu_k \, = \, 2 \left( \frac{k}{\Lambda} \right)^{1/2} \sqrt{\dot \phi}_e \, = \,
2 \left( \frac{2}{3}\right)^{1/8} \left( \frac{k}{\Lambda} \right)^{1/2}  m_{pl}^{1/4} \mu^{3/4} \, .
\ee
The factor $F_k$ by which the amplitude of $A_k$ is amplified is
\be
F_k \, = \, e^{X_k} \, ,
\ee
with
\be
X_k \, = \, \frac{1}{4} T \mu_k \, ,
\ee
where $T$ is the period. The quarter period is given by
\be
\frac{1}{4} T \, = \, \frac{\phi_e}{{\dot \phi}_e} \, .
\ee
Combining these equations yields
\be
X_k \, = \, X_{k_*} \bigl( \frac{k}{k_*} \bigr)^{1/2}  ,
\ee
with
\be \label{ampl2}
X_{k_*} \, = \, 2 \, \left( \frac{2}{3}\right)^{1/2} \frac{m_{pl}}{\Lambda} \, .
\ee
Comparing the amplification factors $F_I$ and $F_k$ (see (\ref{ampl1}) and (\ref{ampl2})
one sees that at the value $k = k_*$ they have similar magnitudes.

The mode function after one period of oscillation of $\phi$ is thus given by
\be
\label{modefunction1}
A_{k+}  \, = \, 
\frac{2^{-1/4}}{\sqrt{2k}} e^{X_k} \left( \frac{k}{ \xi a H}\right)^{1/4} e^{\pi \xi -  4 \xi \sqrt{ k / 2 \xi a H} }\, .
\ee
As long as the expansion of the universe can be neglected, and before back-reaction shuts off
the resonance, the gauge field fluctuations grow by the same factor in each period. Hence,
after $N$ periods we obtain
\be
\label{modefunction1}
A_{k+}   \, = \, 
\frac{2^{-1/4}}{\sqrt{2k}} e^{N X_k} \left( \frac{k}{ \xi a H}\right)^{1/4} e^{\pi \xi - 4 \xi \sqrt{ k /2 \xi a H}} \, .
\ee
Comparing the expressions for the period $T$ and the Hubble expansion rate $H$ at the
end of inflation we see that right at the end of inflation $T \sim H^{-1}$ and hence the
expansion of space cannot be neglected. However, once reheating starts, $\phi$ decreases
and hence $T$ decreases and the expansion of space becomes negligible. The Floquet
exponent can be taken to be approximately constant during half of each period, and vanishing
for the other half. Hence, over a period $(0, t)$ of reheating, the increase in the amplitude is
\be
F_k \, \sim e^{\frac{1}{2} \mu_k t} \, ,
\ee
and the gauge field amplitude becomes
\be
\label{modefunction}
A_{k+}   \, = \, 
\frac{2^{-1/4}}{\sqrt{2k}} e^{\frac{1}{2} \mu_k t}  \left( \frac{k}{ \xi a H}\right)^{1/4} e^{\pi \xi - 2\sqrt{2} \xi \sqrt{ k / \xi a H}} \, .
\ee
There is also an amplification for the $(-)$ polarization, $A_{k-}$, but this mode is suppressed during inflation, and enters preheating with a different mode function.

\section{Gauge Field Energy Density Fluctuations}

We have thus far computed the gauge fields produced during preheating. This sources an energy density 
perturbation, $\delta \rho_A$, which we will now focus on. The gauge field energy density is defined as 
(in (-,+,+,+) signature)
\begin{equation}
\rho_A (x,t) \, = \, - T^{0} _0  \, ,
\end{equation}
where $T_{\mu \nu}$ is given by (again in (-,+,+,+) signature, and assuming a Lagrangian 
$\mathcal{L}=(1/4)F^2$),
\begin{equation}
T_{\mu \nu} \, = \, - \frac{1}{4}g_{\mu \nu} F^2 +  F_{\mu \lambda} F^{\lambda} _\nu \, .
\end{equation}
In terms of the gauge field $A_\mu$, and without any gauge fixing, this reduces to
\begin{equation}
\rho_A (x,t) \, = \, -\frac{1}{2} (\partial^0 A^i - \partial^i A^0)(\partial_0 A_i - \partial_i A_0) + \frac{1}{4}(\partial^i A^j - \partial^j A^i)(\partial_i A_j - \partial_j A_i) \, .
\end{equation}
We can fix the gauge by setting $A_0 =0$. The leading term on cosmological scales is given by
\begin{equation}
\rho_A(x,t) \, \simeq \,  -\frac{1}{2} \partial^0 A^i  \partial_0 A_i \, .
\end{equation}

To find the Fourier modes of $\rho_A(x,t)$, we first expand $A_{\mu}$ in terms of \emph{classical} oscillators
\begin{equation}
A_\mu (x,t) \, =  \, \displaystyle \sum_{\lambda=+,-} \int \frac{\mathrm{d}^3 k}{(2 \pi)^3} \left[ \epsilon _\mu ^\lambda A_\lambda (k,t) \alpha_k e^{i k x} + {\epsilon _\mu ^{\lambda}}^* A_\lambda (k,t) \alpha^\dagger _k e^{- i k x}\right] \, ,
\end{equation}
where $\alpha_k$ are classical oscillators drawn from a nearly Gaussian distribution, satisfying 
\be
\langle \alpha_k \alpha_{k'}  \rangle \, = \, (2 \pi)^3 \delta^3 (k+k') \, ,
\ee
where the angular brackets stand for ensemble averaging.
We can expand $\rho$ in a similar fashion
\begin{equation}
\rho_A (x,t) \, =  \, \displaystyle  \int \frac{\mathrm{d}^3 k}{(2 \pi)^3} \rho_{Ak} \beta_k e^{i k x} + c.c. \, ,
\end{equation}
where $\beta_k$ are a different set of classical oscillators, whose distribution function can be determined in 
terms of the $\alpha_k$. The Fourier modes of $\rho(x,t)$ are simply a convolution of Fourier modes of the 
gauge field $A_{\mu}$
\begin{equation}
\label{rhoAkoscillatorexpansion}
\rho_{Ak} \beta_k \, = \, + \frac{1}{2} a^{-2} \int \frac{\mathrm{d}^3{k'} }{(2 \pi^3)} \dot{A}_{k'+}  \dot{A}_{(k-k')+} \alpha_{k'} \alpha_{k-k'}  \, ,
\end{equation}
where the mode function $A_k$ is given by equation (\ref{modefunction}). There is a gradient term $k^2 A_k ^2$ which is comparable in magnitude to the time-derivative term, and thus changes $\rho_{Ak}$ by a factor of two.

We can use the above expression to straightforwardly calculate the background energy density 
in the gauge field and the spectrum of the gauge fluctuations. The homogenous background 
energy density is simply $\langle \rho_A(x,t) \rangle$, and we define the fluctuations $\delta \rho_A(x,t)$ 
about this background as $\delta \rho_A= \rho_A - \langle \rho_A \rangle$, such that 
$\langle \delta \rho_A \rangle =0$, and the variance of fluctuations is simply 
$\langle \delta \rho_A ^2 \rangle = \langle \rho_A ^2 \rangle - \langle \rho_A \rangle^2$. A simple 
calculation shows that the background is given by
\begin{equation}
\label{rhoAmodefunctionintegral}
 \langle \rho_A(x,t) \rangle \, = \, \frac{1}{2} a^{-2} \int \mathrm{d}^3{k}  |\dot{A}_{k+}|^2 \, .
\end{equation}
The dominant contribution to the integral comes from the maximally amplified mode $k=k_*$, and 
we can hence approximate it as
\begin{equation}
\label{backgroundrhoA}
\langle \rho_A (x,t) \rangle \, \sim \,  \sqrt{2}\, a^{-2} \, e^{2 \mu_* t} (\mu_* k_*)^2   
e^{- 2 \sqrt{2}} \cdot e^{2 \pi \xi} \, ,
\end{equation}
where $\mu_* \equiv \mu_{k_*}$
From this we see that  the amplitude of $\langle \rho \rangle$ depends inversely on the 
UV scale $\Lambda$, since a smaller $\Lambda$ means an increased $k_*$. 

The mode function of fluctuations can be straightforwardly computed using the definition 
$\delta \rho_A= \rho_A - \langle \rho_A \rangle$ in conjunction with equation 
(\ref{rhoAkoscillatorexpansion}) and the approximation that the $\beta_k$ are drawn from a 
nearly Gaussian distribution, i.e. $\langle \beta_k \beta_{k'} \rangle =(2 \pi)^3 \delta^3 (k+k')$. 
The exact $\beta_k$ are \emph{not} drawn from a Gaussian distribution, but as we have the 
modest goal of computing power spectra (i.e. two-point statistics), this is not an important 
distinction. The dominant term in $\delta \rho_{Ak}$ is
\begin{equation}
\label{rhoAkSolved}
|\delta \rho_{Ak}|^2 \, \simeq \, \frac{1}{4} \,a^{-4}  \int \mathrm{d}^3 q \, |\dot{A}_{q} |^2 |\dot{A}_{k-q} |^2 \, .
\end{equation}
For modes in the IR, i.e. $k \ll k_*$, this integral is highly peaked at $q = k_*$ and we can 
find
\begin{equation}
\label{rhoAkSolvedIR}
|\delta \rho_{Ak}|^2 \, \simeq \,  \frac{\langle \rho_A \rangle^2}{ k_* ^3} \, .
\end{equation}
Note, in particular, that the resulting power spectrum of gauge field fluctuations is
highly blue. The spectral index is $n_s = 4$.

\section{Back-Reaction Considerations}

The exponential increase in the gauge field value cannot continue
forever. Eventually, the tachyonic resonance will be shut off by
back-reaction effects. Back-reaction in a two field toy model of
parametric resonance was considered in \cite{RHB_BR}, where it
was concluded that back-reaction does not prevent the exponential
production of entropy fluctuations before these perturbations become
important. In this subsection we estimate how long the tachyonic
resonance in our model will last until back-reaction becomes important.

We will consider the two most important back-reaction effects
involving gauge field production.
The first is the effect of gauge field production on the axion field
dynamics, the dynamics driving the instability. Recall that the axion equation 
of motion is given by
\begin{equation} \label{scalarEoM}
\ddot{\phi} + 3 H \dot{\phi} + V_{,\phi} \, = \,  \frac{1}{\Lambda} \langle F \tilde{F} \rangle,
\end{equation}
where $\langle  F \tilde{F} \rangle$ refers to enseble or spatial averaging as was done to 
determine $\langle \rho_A \rangle$ in the previous section. To obtain
an order of magnitude estimate of when back-reaction becomes important,
we can compare the term on the right hand side of (\ref{scalarEoM}) with the
force driving the oscillations. The first condition of `small backreaction' comes
from demanding that the force term dominates. This translates to
\begin{equation}
\label{eq:condition1}
\langle V_{,\phi} \rangle_{rms} \gg \langle \frac{F \tilde{F}}{\Lambda} \rangle_{RMS} \, .
\end{equation}
We can estimate the order of magnitude of the right-hand side of the above
equation by $\Lambda^{-1} \rho_A$, and hence the condition (\ref{eq:condition1})
becomes
\be \label{BRcond1}
V^{\prime} \, \gg \, \frac{1}{\Lambda} \rho_A \, .
\ee

The second back-reaction condition comes from demanding that the energy density
is dominated by the scalar field, i.e.
\be \label{BRcond2}
V \, \gg \, \rho_A \, .
\ee
In this equation, the value of $\phi$ appears. We will use the value at the end
of inflation.

For the axion monodromy potential we are using, the two conditions differ by a
factor $\Lambda / m_{pl}$. Inserting the expression (\ref{backgroundrhoA})
into the first back-reaction criterium (\ref{BRcond1}) yields
\be
2 \mu_* t \, = \, - 2 \pi \xi + 3\, {\rm ln}\left(\frac{\Lambda}{\mu}\right) + 2\, {\rm ln}\left(\frac{\Lambda}{m_{pl}} \right)
\ee
for the time interval $t$ before back-reaction becomes important, whereas the
second condition (\ref{BRcond2}) yields
\be
2 \mu_* t \, = \, - 2 \pi \xi + 3\, {\rm ln}\left(\frac{\Lambda}{\mu} \right) + 3 \, {\rm ln}\left(\frac{\Lambda}{m_{pl}} \right)
\ee
which is a stronger condition if $\Lambda < m_{pl}$ and weaker otherwise.

The amplitude of the gauge field energy density fluctuations when back-reaction
becomes important then is bounded from above by 
\bea \label{rhoAfluct}
\delta \rho_{Ak} \, &\sim& \, \frac{V}{ k_*^{3/2}} \,\,\,\,\,\,\,\,  {\rm for} \,\,\, \Lambda > m_{pl} \,  \\
\delta \rho_{Ak} \, &\sim& \, \frac{V}{ k_*^{3/2}} \frac{\Lambda}{m_{pl}} \,\,\,\, {\rm for} \,\,\,
\Lambda < m_{pl} \nonumber \, . 
\eea
Note that there can be back-reaction effects from the production of other
fields which may turn off the resonance much earlier. Since we are interested
in obtaining upper bounds on the effects generated by gauge field production,
we will work with the above upper bounds.
 
\section{Induced Curvature Perturbations}

During reheating purely adiabatic fluctuations on super-Hubble scales cannot be
amplified since it can be shown that the curvature fluctuation variable $\zeta$
is conserved. This can be shown in linear cosmological perturbation theory 
\cite{BST, BK, Lyth, FB}, but the result holds more generally (see e.g. \cite{Niayesh, LV}).
On the other hand, entropy fluctuations can be parametrically amplified during
reheating \cite{BV, FB2} (see also \cite{Kaiser}). Entropy fluctuations inevitably
seed a growing curvature perturbation. Thus, in the presence of entropy modes
it is possible to obtain an exponentially growing curvature fluctuation
on super-Hubble scales (see e.g. \cite{Keshav} for some studies of this
question in earlier string-motivated models of inflation).
 
Consider $\zeta$, the curvature perturbation on uniform density hypersurfaces. This
is the variable which determines the amplitude of the CMB anisotropies at late
times (see \cite{MFB} for a detailed overview of the theory of cosmological
perturbations). In the absence of entropy fluctuations, this variable is conserved
on super-Hubble scales \cite{BST, BK, Lyth, LV}. However, in the presence of entropy
perturbations, a growing mode of $\zeta$ is induced on super-Hubble
scales, as already discussed in the classic review articles on cosmological
perturbations \cite{SasKod, MFB} and as applied to axion inflation in \cite{ABT}. 
For more modern discussions the reader is referred to \cite{Gordon, MW}.  The equation 
of motion for $\zeta_k$ ($k$ denotes the comoving wavenumber)
is given by Equation 3.29 of \cite{MW}
\begin{equation}
\label{EOMzeta}
\dot{\zeta_k} \, =  \, - \frac{H}{p+\rho} \delta P_{nad, k} + 
\frac{1}{3H} \frac{k^2}{a^2} \left( \Psi_k - \zeta_k \right) + \frac{k^4}{9 H \dot{H}} \Psi_k \, ,
\end{equation}
where $\Psi_k$ is the gauge invariant curvature perturbation in longitudinal gauge,
$p$ and $\rho$ are the total pressure and energy densities, respectively and
$\delta P_{nab, k}$ is the non-adiabatic pressure perturbation
On large length scales the dependence on $\Psi$ disappears and the evolution 
equation is simply
\begin{equation}
\label{zetadot}
\dot{\zeta} \, = \, - \frac{H}{p+\rho} \delta P_{nad} \, .
\end{equation}
Note that $\zeta$ is dimensionless. 

The non-adiabatic pressure perturbation $\delta P_{nad}$ is the sum of an intrinsic 
and a relative perturbation
\begin{equation}
\delta P_{nad} \, = \, \delta P_{int} + \delta P_{rel} \, .
\end{equation}
The intrinsic non-adiabatic pressure perturbation is the sum
\begin{equation}
\delta P_{int} \, = \,  \sum_{\alpha} \delta P_{int, \alpha} \, = \, 
\sum_{\alpha} \left( \delta p_{\alpha} - c_\alpha ^2 \delta \rho_\alpha \right) \, ,
\end{equation}
while the relative non-adiabatic pressure perturbation is given by
\begin{equation}
\label{Prel}
\delta P_{rel} \, = \, - \frac{1}{6 H \dot{\rho}} 
\displaystyle \sum _{\alpha \beta} \dot{\rho}_{\alpha} \dot{\rho}_{\beta} (c_\alpha ^2  - c_{\beta}^2) S_{\alpha \beta}  \, ,
\end{equation}
where $S_{\alpha \beta}$ is the relative entropy perturbation
\begin{equation}
\label{RelativeEntropyPert}
S_{\alpha \beta} \, = \, - 3 H \left( \frac{\delta \rho_\alpha}{\dot{\rho_\alpha}} - \frac{\delta \rho_\beta}{\dot{\rho}_\beta} \right) \, .
\end{equation}
In the above equations, the sum runs over the different components of matter, and $c_{\alpha}^2$ is
the square of the speed of sound of the $\alpha$ component of matter.

The above set of equations can be rewritten in a more compact form (see e.g. \cite{Liddle})
\begin{equation}
\label{pnad}
\delta P_{nad} \, = \, \dot{p} \left( \frac{\delta p}{\dot{p}} - \frac{\delta \rho}{\dot{\rho}} \right) \, ,
\end{equation}
where in our case the total pressure is the sum of the contributions from the $\phi$
field and from the gauge field, i.e. $p = p_{\phi}+ p_{A}$, and similarly for $\rho$, and 
we have set the intrinsic entropy perturbations to zero. For a background that is 
dominated by $\phi$, and with $\delta \rho_A > \delta \rho_\phi$, the above 
non-adiabatic pressure perturbation is simply
\begin{equation}
\delta P_{nad} \, = \, \dot{p_\phi} \left( \frac{\delta p_A}{\dot{p}_\phi} - \frac{\delta \rho_A}{\dot{\rho}_\phi} \right) \, ,
\end{equation}
and the evolution equation of $\zeta$ is given by
\begin{equation}
\label{zetasimple}
\dot{\zeta} \, = \,  - \frac{H }{\rho_\phi + p_{\phi}} \left( \frac{1}{3} - c_{s\phi} ^2 \right) \delta \rho_A \, .
\end{equation}

In our case, the gauge field energy density fluctuations $\delta \rho_A$ is increasing 
exponentially with a Floquet exponent $2 \mu_*$ during the preheating phase, as shown
in earlier sections. Hence, integrating over time, we get
\begin{equation}
\label{zetasimple2}
\Delta \zeta_k \, = \,  -  \mu_*^{-1} \frac{H }{\rho_\phi + p_{\phi}} \left( \frac{1}{3} - c_{s\phi} ^2 \right) \delta \rho_{Ak} \, ,
\end{equation}
where the wavenumber $k$ and the density fluctuation $\delta \rho_A$ are Fourier
space quantities. However, since it follows from Section V that $\delta \rho_A$ is
independent of $k$, we find that the power spectrum of the induced fluctuations of
$\zeta$ is
\be
{\cal P}_{\Delta \zeta}(k) \, \sim \, k^3 \, ,
\ee
which corresponds to a highly blue tilted spectrum with index $n_s = 4$. Since the
spectrum has such a large blue tilt, there are no constraints on our model coming
from demanding that the induced curvature fluctuations do not exceed the observational
upper bounds.

\section{Primordial Black Hole Constraints}

Since the power spectrum of induced curvature fluctuations is highly blue,
we have to worry about the possible constraints on the model coming from
over-production of primordial black holes. Primodial black holes are constrained
by a set of cosmological observations, beginning with the original constraints
coming from the observational bounds on cosmic rays produced by radiating
black holes \cite{original}. Primordial black hole production during reheating
has been considered in simple two field inflation models in \cite{GM}, and in
models with spectra with a distinguished scale in \cite{Blais}.

In the context of an inflationary cosmology, primordial black holes of mass $M$
can form when the length scale associated with this mass (i.e. the length $l$ for
which the mass inside a sphere of radius $l$ equals $M$) enters the Hubble
radius. The number density of black holes of this mass will depend on the
amplitude of the primordial power spectrum \footnote{The are 
numerous subtleties in computing the precise number density, which tend to 
suppress the number of primordial black holes formed, see e.g. \cite{Kuhnel:2015vtw} 
and references therein. These details will not be important for our analysis.}.

Since in our case the power spectrum is highly blue, the tightest constraints
will come from the smallest mass for which cosmological constraints exist.
These correspond to black holes with a mass such that they evaporate during
nucleosynthesis. The extra radiation from these black holes would act as
an extra species of radiation, and would destroy the agreement between the
theory of nucleosynthesis and observations (see \cite{BHreviews} for
reviews). The smallest length scale (i.e. largest wavenumber $k$) for
which constraints exist is \cite{Green}
\be
k_{max} \, \sim \, 10^{19} {\rm Mpc}^{-1} ,
\ee
and the approximate bound on the power spectrum is
\be \label{constraint}
{\cal P}_{\zeta}(k_{max}) \, < \, 10^{-1.5} \, .
\ee
In fact, the bound for smaller values of $k$ has comparable amplitude.

The power spectrum including the induced curvature perturbations is given by
\begin{equation}
P_{\zeta}(k) \, = \, \frac{k^3}{(2 \pi)^2} | \mathcal{A}_0 k ^{-3/2 } + \Delta \zeta_k|^2 ,
\end{equation} 
where $\mathcal{A}_0 \sim 10^{-10}$ is the amplitude of the power spectrum at the 
pivot scale $k=k_0=0.05 {\rm Mpc}^{-1}$, and we have approximated the spectrum of 
curvature perturbations from inflation to be scale invariant. We already computed 
the value of the induced curvature fluctuations $\Delta \zeta$ in
the previous section in Eq. (\ref{zetasimple2}). Inserting the values from
(\ref{rhoAfluct}), (\ref{kstar}) and (\ref{xieq}) we obtain the following
expressions for the leading order correction to the power spectrum of curvature fluctuations
\bea \label{final}
\Delta {\cal P}_{\zeta}(k) \, &=&   \, {\cal O}(10^{-3}) \, \sqrt{\mathcal{A}_0} \,  \, k^{3/2} \, \frac{\Lambda^{5/2}}{m_{pl}^{7/4} \mu^{9/4}}\,\,\,\, {\rm{for}} \,\,\, \Lambda > m_{pl} \nonumber \\
\Delta {\cal P}_{\zeta}(k) \, &=& \, {\cal O}(10^{-3}) \,  \sqrt{\mathcal{A}_0} \,  \, k^{3/2} \, \frac{\Lambda^{7/2}}{m_{pl}^{11/4} \mu^{9/4}} \,\,\,\, {\rm{for}} \,\,\,\Lambda  < m_{pl} \, ,
\eea 
corresponding to a spectrum with index $n_s=5/2$. These expressions hold if the 
exponential growth of the curvature
fluctuations is only limited by the back-reaction effects studied in Section VI.
Other effects may terminate the growth earlier. Hence, the above
equations provide upper bounds on the amplitude of the induced curvature
perturbations.

For the largest value of $k$ for which the primordial black hole
constraints apply we have
\be
\frac{k}{m_{pl}} \, \sim \, 10^{-39} \, .
\ee
Inserting this value into (\ref{final}) we find that the primordial black hole
constraint (\ref{constraint}) is trivially satisfied for the realistic range of
values of $\Lambda$.

\section{Conclusions}

In this paper we have considered a minimal axion monodromy model and
have calculated the spectrum of the curvature perturbations
induced by the entropy mode associated with the gauge field to which
the axion couples. We find that the leading correction to the curvature spectrum 
is blue with spectral index
$n_s = 5/2$. Hence, there are no constraints from large scale cosmological
observations. On the other hand, since the spectrum is blue, there is
a danger of overproduction of primordial black holes. We find, however,
that the amplitude of the spectrum is too low even on the smallest
scales for which cosmological constraints exist.

Realistic axion monodromy models, on the other hand, typically
contain many other scalar fields which can source entropy modes,
and these modes could, in principle, pose cosmological problems.
It is reassuring, however, that the prototypical minimal axion monodromy
model is safe from the constraints studied in this paper.

\section*{Acknowledgement}
\noindent

Two of the authors (RB and EM) wish to thank the Institute for Theoretical Studies of the ETH
Z\"urich for kind hospitality while a portion of this work was completed. RB acknowledges 
financial support from Dr. Max
R\"ossler, the Walter Haefner Foundation and the ETH Zurich Foundation, and
from a Simons Foundation fellowship. The research at McGill is supported in
part by funds from NSERC and the Canada Research Chair program.
EM is supported in part by a NSERC PGS D fellowship. HBM is supported in part by a 
MSRT fellowship from Iran.

\appendix 

\section{String Theory Model Building Constraints on the Coupling of $\phi$ to $F \tilde{F}$}
\label{sec:coupling}

A realistic universe built from axion monodromy neccessarily has three sectors: 
(1) the inflation sector, (2) the moduli stabilization sector, and (3) the standard model. 
However, without considering all three, there is already interesting couplings purely 
in the inflation sector.

In (the standard story of) axion monodromy inflation \cite{momo}, a potential an axion 
field is induced by the DBI action of a D5 action. The world volume action of the D5-brane 
receives corrections at each order in the string coupling constant $\alpha'$. In particular, 
the Chern-Simons part of the action has an $\alpha'^2$ correction:
\begin{equation}
\delta S_{CS} \, = \, - \mu_5 (2 \pi \alpha')^2 \int C_0 B_2 \wedge F \wedge F \, ,
\end{equation}
where $C_0$ is the RR $0$-form potential, $B_2$ is the NSNS two-form, $F$ is the world volume 
gauge field (see, for example, Equation 2.6 of \cite{Shiu}). There are also corrections to the 
DBI action, but the coupling of these corrections to the two-form $B_2$ is not known 
(see for example equation (2.4) of \cite{Shiu}).

From the above action we can derive the 4d interaction,
\begin{equation}
\mathcal{L}_{int} \, = \, \frac{1}{\Lambda}\phi F \tilde{F} \, ,
\end{equation}
where $\Lambda$ is given by
\begin{equation}
\frac{1}{\Lambda} \, \equiv \, 2 \mu^3 l^3  (2 \pi \alpha')^2 C_0   \, ,
\end{equation}
where $\mu^3$ is the coupling constant appearing in the axion potential, and $l^2$ is
the size of $\Sigma_2$ (the two-cycle wrapped by the brane) in string units.The 
constants in $\Lambda$ are constrained by the consistency 
of the model. We can write $\alpha'$ in terms of the string 
mass scale, $M_{s} = 1/\sqrt{2 \pi \alpha'}$, such that the coupling takes the form
\begin{equation}
\frac{1}{\Lambda} \, = \, 2 l^3 C_0 \left( \frac{\mu}{M_s}\right)^3 \frac{1}{M_s} \, .
\end{equation}

Let us consider the size of the parameters. Firstly, moduli stabilization requires a stack of 
D7 branes. These branes source $C_0$, and the value of $C_0$ (which is dimenionless) is 
roughly equal to the number of D7 branes. For more details on the supergravity background 
in the presence of a stack of D7 branes, see e.g. \cite{Ouyang}. In realistic models of 4d 
physics (see e.g. \cite{Rummel:2011cd})
\begin{equation}
C_0 \, \sim \, 10^2 \, .
\end{equation}

Secondly, the value of $\mu$ is chosen such that the amplitude of cosmological
perturbations arising from axion monodromy inflation matches
observations. The potential is
\begin{equation}
V_{int} \, = \, \mu^3 \phi \, ,
\end{equation}
where $\mu$ is given by
\begin{equation}
\mu \, = \, \frac{\mu_5 \alpha'}{f g_s} \, ,
\end{equation}
where $f$ is the axion decay constant. Note that the axion decay constant enters the 
potential and the interaction term with the same power, and thus it is impossible to change 
the relative strength of the interaction by fine-tuning the axion decay constant. Consistency 
with observations requires
\begin{equation}
\mu^3 \, = \, \left( 6 \times 10^{-4} m_{pl} \right)^3 \, .
\end{equation}

Thirdly, the internal space must be of an `intermediate' size: if the internal space is too 
small then the supergravity approximation (and the DBI action) ceases to be the correct 
description of physics, while if the internal space is too large then the 4d Newton's 
constant is too small. A reasonable value of the volume of the internal space is 
$Vol(X_6) \sim 10^6$ in units of $\alpha'$, which corresponds to a length scale $l_{X_{6}} \sim 10$.

These constraints determine to a great degree the allowed value of the parameter 
$\Lambda$. A consistent value is:
\begin{equation}
\frac{1}{\Lambda} \, \sim \, 10^{-5} \left(\frac{m_{pl}}{M_{s}}\right)^4 \frac{1}{m_{pl}} \, ,
\end{equation}
where we took $C_0 = 50$, $l=10$, and $\mu = 6 \times 10^{-4} m_{pl}$.

\end{document}